\documentstyle[12pt]{article}
\newcommand{\beq}{\begin{equation}}
\newcommand{\eeq}{\end{equation}}

\newcommand{\bra}{\begin{array}}

\newcommand{\era}{\end{array}}

\newcommand{\al}{\alpha}

\newcommand{\ga}{\gamma}

\newcommand{\de}{\delta}

\newcommand{\Te}{\Theta}

\newcommand{\ot}{\otimes}

\newcommand{\ep}{\epsilon}

\begin{document}
\begin{center}

{\bf\huge Fractional Supersymmetry through Generalized Anyonic algebra}\\

\vskip1.2truecm
{\bf J. Douari}
\footnote[1]
{Junior Associate of the Abdus Salam ICTP.\\
\null\hspace{0.5cm} E-mail: douarij@ictp.trieste.it}\\
\small{The Abdus Salam ICTP - strada costiera 11, 34100 Trieste, Italy.}\\
\small{and}\\
\small{Facult\'e des Sciences, D\'epartement de Physique, LPT-ICAC}\\
\small{Av. Ibn Battouta, B.P.1014, Agdal, Rabat, MOROCCO}\\
\vspace*{0.5cm}
{\bf Y. Hassouni}
\footnote[2] 
{Regular Associate of the Abdus Salam ICTP.\\
\null\hspace{0.5cm} E-mail: Y-hassou@fsr.ac.ma}\\
\small{Facult\'e des Sciences, D\'epartement de Physique, LPT-ICAC}\\
\small{Av. Ibn Battouta, B.P.1014, Agdal, Rabat, MOROCCO}

\end{center}
\hoffset=-1cm\textwidth=20cm\pagestyle{empty}
\vspace*{0.5cm}
\section*{Abstract}
\hspace{.3in}The construction of anyonic operators and algebra is
generalized by using quons operators. Therefore, the particular version
of fractional supersymmetry is constructed on the two-dimensional
lattice by associating two generalized anyons of different kinds. The
fractional supersymmetry Hamiltonian operator is obtained on the
two-dimensional lattice and the quantum algebra $U_{q}(sl_{2})$ is
realized.
\newpage
\section{Introduction}
\hspace{.3in}A grown interest has been devoted, recently, to fractional
statistics$^{\cite{1,2}}$ describing particles with fractional
spin$^{\cite{3,4,5}}$. These particles are called anyons and interpolate
between bosons and fermions, it has been shown also that anyons are
non-local particles defined on the two-dimensional space.
Mathematically, the group symmetry associated to the anyonic systems
involves some special Lie algebras appearing as a quantum deformation of
the usual Lie algebras and Lie groups$^{\cite{6,7,8,9,10}}$. Indeed, one
can prove that quantum groups$^{\cite{11}}$ are the mathematical objects
allowing the description of these particular systems$^{\cite{12,13}}$.
Many works have been devoted, in the last few years, to the construction
and the realization of quantum groups. We note for example the anyonic
algebra$^{\cite{14}}$ leading to the obtaining of these quantum
algebras. The anyonic operators are introduced on the two-dimensional
lattice as a non-local operators seen as a generalization of generators
of the Jordan-Wigner transformation$^{\cite{15}}$.\\ \hspace*{.3in}The
present work concerns the study on the two-dimensional lattice of N=2
fractional supersymmetry (FSUSY)$^{\cite{16,17,18,19}}$. We obtain
exactly the same algebraic structure of this latter starting from the
introduction of the generalized anyonic algebra. Then we introduce some
special operators which can be seen as a generalized anyonic operators.
We notice that this construction is different from the one used to
reproduce the N=2 FSUSY algebra basing on the quonic
algebra$^{\cite{19}}$. This algebraic study allows us to find the
Hamiltonian operator describing one N=2 FSUSY system.\\
\hspace*{.3in}This paper is organized as follows : In section $2$, as a
generalization, we construct new exotic operators in using quonic
operators as basic ones. Therefore we realize the generalized anyonic
operators and the corresponding algebra. In the third section, the last
new operators and algebra will be used to construct N=2 FSUSY on the
two-dimensional lattice corresponding to anyonic systems as a certain
coupling of two generalized anyonic oscillators of different kinds
($\ga$ and $\de$)$^{\cite{14}}$. We proceed as in the work \cite{19}
where we have constructed the N=2 FSUSY through two different quons.
Another result consists on the realization of the quantum algebra
$U_{q}(sl_{2})$ using the supercharges already constructed. Finally,
section $4$ presents some concluding remarks.
\section{Generalized anyonic algebra}
\hspace{.3in}Let us recall at first the anyonic operators which are seen
as non-local two-dimensional operators interpolating between bosonic and
fermionic ones. The generalization of these operators allows us to
introduce the generalized anyons. We will use the famous angle function
$\Te(x,y)$ appearing in the construction and the description of anyons
in the work \cite{6}.\\ \hspace*{.3in}We start by giving a brief review
on this angle function. One designs by $\ga_{x}$ the cut associated to
each point $x$ that we denote by $x_{\ga}$ on the two-dimensional
lattice $\Omega$. Denoting by $\Omega^{*}$ the dual lattice of $\Omega$;
it is a set of points $x^{*}=x+0^{*}$ with
$0^{*}=(\frac{1}{2}\ep,\frac{1}{2}\ep)$ the origin of $\Omega^{*}$ and
$\ep$ its spacing which eventually will be sent to zero. In this case
$\ga_{x}$ will be on $\Omega^{*}$ from minus infinity to $x^{*}$ along
the $x$-axis. One considers another type of cuts. We choose the set of
cuts $\de_{x}$ coming from plus infinity to $^{*}x=x-0^{*}$ along the
$x$-axis. Consequently, the two types of cuts $\ga_{x}$ and $\de_{x}$
involve an ordering and opposite ordering respectively of points $x$ on
$\Omega$. This is described by the following proposition
\beq
x_{\de}<y_{\de} \Leftrightarrow  x_{\ga}>y_{\ga} \Leftrightarrow x>y \Leftrightarrow \left\{
\bra{cc}
x_{2} > y_{2},& \\
x_{1}>y_{1}, x_{2}=y_{2}
\era
\right.
\eeq
Owing to the equation $(1)$ the angle functions satisfy
\beq
\bra{rcl}
\Te_{\ga_{x}}(x,y)-\Te_{\ga_{y}}(y,x)&=&
\left\{\bra{rcl}
\mbox{$\pi$,  $x>y$} \\
\mbox{$-\pi$, $x<y$}
\era 
\right.
\\
\Te_{\de_{x}}(x,y)-\Te_{\de_{y}}(y,x)&=&
\left\{\bra{rcl}
\mbox{$-\pi$,  $x>y$} \\
\mbox{$\pi$, $x<y$}
\era
\right.
 \\
\Te_{\de_{x}}(x,y)-\Te_{\ga_{x}}(x,y)&=&
\left\{\bra{rcl}
\mbox{$-\pi$,  $x>y$} \\
\mbox{$\pi$, $x<y$}
\era
\right.
 \\
\Te_{\de_{x}}(x,y)-\Te_{\ga_{y}}(y,x)&=&0, \forall x,y\in\Omega.
\era
\eeq
\hspace*{.3in}Now let us introduce the operators $K_{i}(x_{\al})$ those
called the disorder ones. They are expressed by
\beq
K_{i}(x_{\al})=e^{i\nu\sum\limits_{y\neq x}\Te_{\al_{x}}(x,y)[N_{i}(y)-\frac{1}{2}]}
\eeq
with $\al_{x}=\ga_{x}$ or $\de_{x}$ and $i=1,2,...,n, n\in\bf{N}\rm$. In the equality $(3)$ $N_{i}(y)$ is nothing but the number operator of quons on the two-dimensional lattice, defined by the generators $a_{i}^{\dag}(x)$ and $a_{i}(x)$ as follows
\beq
\bra{rcl}
a_{i}^{\dag}(x)a_{i}(x)&=&[N_{i}(x)]_{q_{i}}\\
a_{i}(x)a_{i}^{\dagger}(x)&=&[N_{i}(x)+\bf{1}]_{\rm{q_{i}}}
\era
\eeq
where $[x]_{q}=\frac{q^{x}-1}{q-1}$. The operators $a_{i}^{\dag}(x)$ and $a_{i}(x)$ are respectively the creation and annihilation quonic operators on the two-dimensional lattice $\Omega$, satisfying the relations
\beq
\bra{rcl}
\lbrack a_{i}(x),a_{j}^{\dag}(y)\rbrack_{q_{i}^{\de_{ij}}}&=&\de_{ij}\de(x,y)
\\
\lbrack a_{i}(x),a_{j}(y)\rbrack_{q_{i}^{\de_{ij}}} &=&\bra{cc}
0& \mbox{$\forall x,y$, $\forall i,j$}
\era\\
\lbrack a_{i}^{\dag}(x),a_{j}^{\dag}(y)\rbrack_{q_{i}^{\de_{ij}}} &=&\bra{cc}
0& \mbox{$\forall x,y$, $\forall i,j$}
\era\\
\lbrack N_{i}(x),a_{j}(y)\rbrack &=&-\de_{ij}\de(x,y)a_{i}(x)
\\
\lbrack N_{i}(x),a_{j}^{\dag}(y)\rbrack &=&\de_{ij}\de(x,y)
a_{i}^{\dag}(x)\\
\era
\eeq
The Dirac function is defined by
\beq
\de(x,y)=\left\{\bra{cc}
1& \mbox{if $x=y$} \\
0&\mbox{ if $x\neq 0$}
\era
\right.
\eeq
In the relations $(5)$, one can consider the operators
$a_{i}^{\dag}(x)$, $a_{i}(x)$ and $N_{i}(x)$; those are generators of
the oscillator algebra describing a system of quons. They are seen as a
non-local particles and we define them to satisfy the equality
\beq
(a_{i}(x))^{d_{i}}=(a_{i}^{\dag}(x))^{d_{i}}=0
\eeq
where we take the deformation parameter $q_{i}$ to be a root of unity,
$q_{i}^{d_{i}}=1$ then $q_{i}=e^{i\frac{2\pi}{d _{i}}}$. One can show
that the irreducible representation space (Fock space) of the algebra
(relations $(5)$) is given by the set
\beq
F_{i_{x}}=\lbrace\vert n_{i_{x}}>, n_{i_{x}}=0,1,...,d_{i}-1\rbrace
\eeq
where the notation $i_{x}$ means that this Fock space is introduced in
each site of the $\Omega$. the actions of the generators $a_{i}(x)$,
$a_{i}^{\dag}(x)$ and  $N_{i}(x)$ are expressed by the following
equalities
\beq
\bra{rl}
a_{i}^{\dag}(x)\vert n_{i_{x}}>=\vert n_{i_{x}}+1>,& a_{i}^{\dag}(x)\vert d_{i}-1>=0 \\
a_{i}(x)\vert n_{i_{x}}>=[n_{i_{x}}]_{q_{i}}\vert n_{i_{x}}-1>,& a_{i}(x)\vert 0>=0.
\era
\eeq
\hspace*{.3in}Now we introduce an operators allowing the description of
the N=2 FSUSY on the two-dimensional lattice. This realization is seen
as an anyonic one, in the sense that the generators will depend on a
function which seems to be the angle function used by the authors in the
work \cite{14} when they describe the system of anyons. This leads to
the definition of the operators
\beq
A_{i}(x_{\al})=K_{i}(x_{\al})a_{i}(x),
\eeq
where $K_{i}(x_{\al})$ is the disorder operator introduced in the
equation $(3)$.\\ 
\hspace*{.3in}One proves that these operators obey the
following commutation relations
\beq
\bra{rcl}
\lbrack A_{i}(x_{\ga}),A_{i}(y_{\ga})\rbrack_{q_{i}p^{-1}}&=&\mbox{$0$, $x>y$}\\
\lbrack A^{\dag}_{i}(x_{\ga}),A^{\dag}_{i}(y_{\ga})\rbrack_{q_{i}p^{-1}}&=&\mbox{$0$, $x>y$}\\
\lbrack A_{i}(x_{\ga}),A^{\dag}_{i}(y_{\ga})\rbrack_{q_{i}p}&=&\mbox{$0$, $x>y$}\\
\lbrack A^{\dag}_{i}(x_{\ga}),A_{i}(y_{\ga})\rbrack_{q_{i}p}&=&\mbox{$0$, $x>y$}\\
\lbrack A_{i}(x_{\ga}),A^{\dag}_{i}(x_{\ga})\rbrack_{q_{i}}&=&1\\
\lbrack A_{i}(x_{\ga}),A_{j}(y_{\ga})\rbrack &=&\mbox{$0$, $\forall i\neq j$, $\forall x,y\in \Omega$}\\
\lbrack A^{\dag}_{i}(x_{\ga}),A^{\dag}_{j}(y_{\ga})\rbrack &=&\mbox{$0$, $\forall i\neq j$, $\forall x,y\in \Omega$}\\
\lbrack A^{\dag}_{i}(x_{\ga}),A_{j}(y_{\ga})\rbrack &=&\mbox{$0$, $\forall i\neq j$, $\forall x,y\in \Omega$}\\
\lbrack A_{i}(x_{\ga}),A^{\dag}_{j}(y_{\ga})\rbrack &=&\mbox{$0$, $\forall i\neq j$, $\forall x,y\in \Omega$}.
\era
\eeq
for the anyonic operators of type $\ga$. We point out that the same results can be obtained for the kind $\de$ in replacing $p$ by $p^{-1}$ in $(11)$. It is obvious thus to get the
commutation relations between the different kinds of generalized anyonic operators, we have so
\beq
\bra{rcl}
\lbrack A_{i}(x_{\de}),A_{j}(y_{\ga})\rbrack_{q_{i}^{\de_{ij}}}&=&0, \mbox{$\forall$ $x,y\in \Omega$}\\
\lbrack A_{i}(x_{\de}),A^{\dag}_{j}(y_{\ga})\rbrack_{q_{i}^{\de_{ij}}}&=&
\de_{ij}\de(x,y)p^{-[\sum\limits_{z<x}-\sum\limits_{z>x}][N_{i}(z)-\frac{1}{2}]}
\era
\eeq
with $p=e^{i\nu\pi}$, $\nu$ is seen as the statistical parameter$^{\cite{1,2}}$.\\
\hspace*{.3in}By considering the generalized anyonic oscillators 
$A_{i}(x_{\al})$ and $A^{\dag}_{i}(x_{\al})$, we can demonstrate the following relations
\beq
(A_{i}(x_{\al}))^{d_{i}}=(A^{\dag}_{i}(x_{\al}))^{d_{i}}=0
\eeq
with $\al=\ga, \de$ and $i=1,2,...,N$. The equation $(13)$ can be seen as a generalization of the hard-core condition found when we have study the generalized statistics in the works \cite{21,22}. Indeed, for $d_{i}=2$ one recover the result allowing us, in this work, to think  about some connexion between the generalized statistics and the anyonic ones. Returning to the present paper, the relation $(13)$ can be seen a nilpotency condition leading to the study of quons on the two-dimensional lattice. This equality is obtained starting from the function $K_{i}(x_{\al})$ discussed in the work \cite{14}, which is the subject of the following equations 
\beq
\bra{rcl}
K^{\dag}_{i}(x_{\al})a_{i}(y)&=&e^{i\nu \Te_{\al_{x}}(x,y)}a_{i}(y)K^{\dag}_{i}(x_{\al}) \\
K^{\dag}_{i}(x_{\al})a_{i}^{\dag}(y)&=&e^{-i\nu \Te_{\al_{x}}(x,y)}a_{i}^{\dag}(y)K^{\dag}_{i}(x_{\al}) \\
K_{i}(x_{\al})a_{i}(y)&=&e^{-i\nu \Te_{\al_{x}}(x,y)}a_{i}(y)K_{i}(x_{\al}) \\
K_{i}(x_{\al})a_{i}^{\dag}(y)&=&e^{i\nu \Te_{\al_{x}}(x,y)}a_{i}^{\dag}(y)K_{i}(x_{\al}) \\
K^{\dag}_{i}(x_{\al})K_{i}(y_{\al})&=&K_{i}(y_{\al})K^{\dag}_{i}(x_{\al})
\era
\eeq
\hspace*{.3in}It is proved that, on the above Fock space, these algebraic relations are coherent with the following equalities
\beq
\bra{rcl}
A^{\dag}_{i}(x_{\al})\vert n_{i_{x}}>&=&e^{i\frac{\nu}{2}\sum\limits_{y\neq x} \Te_{\al_{x}}(x,y)}\vert n_{i_{x}}+1> \\
A_{i}(x_{\al})\vert n_{i_{x}}>&=&[n_{i_{x}}]_{q_{i}}e^{-i\frac{\nu}{2}\sum\limits_{y\neq x} \Te_{\al_{x}}(x,y)}\vert n_{i_{x}}-1> \\
A^{\dag}_{i}(x_{\al})\vert d_{i}-1>&=&0 \\
A_{i}(x_{\al})\vert 0>&=&0.
\era
\eeq
We note that the operators $A^{\dag}_{i}(x_{\al})$ and $A_{i}(x_{\al})$ are seen, owing to the relations $(14)$ and $(15)$, as generalized creation and annihilation anyonic operators respectively.
\section{N=2 FSUSY on the two-dimensional lattice}
\hspace{.3in}The construction of generalized anyonic operators allows us to realize the N=2 FSUSY. This realization involves two different generalized of anyons and is analogous to the one obtained via two different quons in the work \cite{19}. We introduce then the fractional supercharges on $\Omega$ as
\beq
\bra{rcl}
Q_{+}(x)&=&A^{\dag}_{1}(x_{\ga})A_{2}(x_{\de})\\
Q_{-}(x)&=&A_{1}(x_{\de})A^{\dag}_{2}(x_{\ga}).
\era
\eeq
basing on the equality $(13)$; one can get
\beq
(Q_{\pm}(x))^{d_{2}}=0,
\eeq
where $d_{2}$ is required to be $<d_{1}$.\\
\hspace*{.3in}In using the construction given in the relation $(16)$, we can get the action of the above fractional supercharges on the Fock space defined by the tensor product
\beq
F_{x}=F_{1_{x}}\ot F_{2_{x}}
\eeq
with $F_{1_{x}}$ and $F_{2_{x}}$ correspond respectively to the different generalized anyons used to introduce the fractional supercharges.\\
\hspace*{.3in}Now we can get easily the relations
\beq
\bra{rcl}
Q_{+}(x)\vert n_{1_{x}}>\ot\vert n_{2_{x}}>=p^{\frac{1}{2}}[n_{2_{x}}]_{q_{2}}\vert n_{1_{x}}+1>\ot\vert n_{2_{x}}-1> \\
Q_{-}(x)\vert n_{1_{x}}>\ot\vert n_{2_{x}}>=p^{\frac{1}{2}}[n_{1_{x}}]_{q_{1}}\vert n_{1_{x}}-1>\ot\vert n_{2_{x}}+1>
\era
\eeq
Through this realization of the fractional supercharges $Q_{\pm}(x)$, we can show that these new operators satisfy the following commutation relation in the case of $x>y$ on $\Omega$
\beq
q_{1}Q_{+}(x)Q_{-}(y)-q_{2}Q_{-}(y)Q_{+}(x)=\de(x,y)P[q_{1}[N_{1}(x)]_{q_{1}}-q_{2}[N_{2}(x)]_{q_{2}}]
\eeq
where the operator $P$ is written as
\beq
P=p^{[\sum\limits_{z<x}-\sum\limits_{z>x}][N_{1}(z)+N_{2}(z)-1]}
\eeq
\hspace*{.3in}One can remark that the equality $(20)$ is not invariant under the
hermitian conjugate. This is related to the fact that these generators involve a
complex numbers $q_{1,2}$ and $p$ which are different from $\pm 1$. To avoid this
difficulty we introduce the hermitian conjugate of the generators $Q_{\pm}(x)$ and after calculation it is easy to obtain the conjugate equation of $(20)$ as 
\beq
q_{1}^{-1}Q_{-}^{\dag}(y)Q_{+}^{\dag}(x)-q_{2}^{-1}Q_{+}^{\dag}(x)Q_{-}^{\dag}(y)=\de(x,y)P^{-1}[q_{1}^{-1}[N_{1}(x)]_{q_{1}^{-1}}-q_{2}^{-1}[N_{2}(x)]_{q_{2}^{-1}}],
\eeq
We can verify also that
\beq
(Q_{\pm}^{\dag}(x))^{d_{2}}=0.
\eeq
for $d_{2}<d_{1}$.\\
\hspace*{.3in}Another reason to introduce this operation is to construct the Hamiltonian operator corresponding to this system. We can thus express the FSUSY Hamiltonian operator as
\beq
\bra{rcl}
q_{1}Q_{+}(x)Q_{-}(y)+q_{1}^{-1}Q_{-}^{\dag}(y)Q_{+}^{\dag}(x)-q_{2}Q_{-}(y)Q_{+}(x)-q_{2}^{-1}Q_{+}^{\dag}(x)Q_{-}^{\dag}(y)=\\
\de(x,y)[Pq_{1}[N_{1}(x)]_{q_{1}}+P^{-1}q_{1}^{-1}[N_{1}(x)]_{q_{1}^{-1}}-Pq_{2}[N_{2}(x)]_{q_{2}}-P^{-1}q_{2}^{-1}[N_{2}(x)]_{q_{2}^{-1}}].
\end{array}
\eeq
Using the relation
\beq
[N_{i}(x)]_{q_{i}^{-1}}=q_{i}^{1-N_{i}(x)}[N_{i}(x)]_{q_{i}}.
\eeq
The RHS of $(24)$ will be rewritten as 
\beq
RHS=\de(x,y)[P^{-1}q_{1}^{-N_{1}(x)}+Pq_{1}][N_{1}(x)]_{q_{1}}-
     [P^{-1}q_{2}^{-N_{2}(x)}+Pq_{2}][N_{2}(x)]_{q_{2}},
\eeq
\hspace*{.3in}The last equation can be interpreted as a Hamiltonian of the generalized anyonic system investigated. So, we can rewrite this Hamiltonian operator as
\beq
H(x)=\sum\limits_{i,j=1,2}\ep_{ij}\frac{\sin[\nu\pi\sum\limits_{z\in\Omega}\aleph(z)+\frac{\pi}{d_{i}}(2N_{i}(x)+1)]-\sin[\nu\pi\sum\limits_{z\in\Omega}\aleph(z)+\frac{\pi}{d_{i}}]}{\sin\frac{\pi}{d_{i}}},
\eeq
where $(\ep_{ij})=\pmatrix{0&1\cr-1&0\cr}$ and $\sum\limits_{z\in\Omega}\aleph(z)=(\sum\limits_{z<x}-\sum\limits_{z>x})(N_{1}(z)+N_{2}(z)-1)$. Up to now, we can recapulate our result which consists on the complet description of the N=2 FSUSY basing on a given anyonic realization.\\
\hspace*{.3in}Furthermore a global version of this realization can be readily constructed as follows
\beq
H=\sum\limits_{x\in\Omega}H(x),
\eeq
where the global supercharges are defined as 
\beq
Q_{\pm}=\sum\limits_{x\in\Omega}Q_{\pm}(x),
\eeq
\hspace*{.3in}In addition, we can link the N=2 FSUSY obtained on the two-dimensional lattice to the "local" algebra $U_{q}(sl_{2})$ in considering $q_{1}=q_{2}=q$. In this particular case we define the three local generators as
\beq
\bra{rcl}
J_{\pm}(x)=P^{-\frac{1}{2}}q^{-\frac{N_{2}(x)}{2}}Q_{\pm}(x)\\
J_{3}(x)=\frac{1}{2}(N_{1}(x)-N_{2}(x)).
\end{array}
\eeq
We can easily check that these local densities of quantum group generators satisfy the following commutation relations
\beq
\bra{rcl}
\lbrack J_{+}(x),J_{-}(y) \rbrack=\de(x,y)[2J_{3}(x)]_{q}\\
\lbrack J_{3}(x),J_{\pm}(y) \rbrack=\pm\de(x,y)J_{\pm}(x).
\end{array}
\eeq
Thus to define the global generators it is sufficient to write
\beq
\bra{rcl}
J_{\pm}=\sum\limits_{x\in\Omega}J_{\pm}(x)\\
J_{3}=\sum\limits_{x\in\Omega}J_{3}(x)
\end{array}
\eeq
and close the $U_{q}(sl_{2})$ algebra as
\beq
\bra{rcl}
\lbrack J_{+},J_{-} \rbrack=[2J_{3}]_{q}\\
\lbrack J_{3},J_{\pm} \rbrack=\pm J_{\pm}.
\end{array}
\eeq
\hspace*{.3in}Consequently, we have close the algebra of $U_{q}(sl_{2})$
generated by $J_{\pm}$ and $J_{3}$ which are built out of generalized
anyonic oscillators.
\section{Conclusion}
\hspace{.3in}To conclude, we can summarize the lines of this paper in
saying that we have constructed a generalized anyonic operators on the
two-dimensional lattice in using the q-bosonic operators or called
quonic operators, as a generalization of ones defined in the paper
\cite{5}. Moreover, we have realized the N=2 FSUSY on the
two-dimensional lattice. Where the supercharges are constructed by
coupling two different generalized anyonic operators, and the FSUSY
Hamiltonian operator of the corresponding system is given. Thus from the
N=2 FSUSY realized the well known algebra $U_{q}(sl_{2})$ is derived.\\
\hspace*{.3in}It will be interesting also to rewrite the obtained FSUSY
Hamiltonian, in the connexion with the gauge theory, as a form of which
we held the term coincidering with the Chern-Simons term. This point is
the matter that we are preparing for the next paper.
\section*{Acknowledgments}
\hspace*{.3in}The author J. Douari would like to thank the
Max-Planck-Institut f\"ur Physik Komplexer Systeme for link hospitality
during the stage in which one part of this paper was done and the
International Atomic Energy Agency and UNESCO for hospitality at the
Abdus Salam International Centre for Theoretical Physics, Trieste. This
work was achieved within the framework of the Associateship Scheme of
the Abdus Salam International Centre for Theoretical Physics. And
special thanks to Professors Manfred Scheunert, Ruibin Zhang and G.
Thompson for an inspiring comment and many useful discussions.

\end{document}